\documentstyle[11pt,moriond,epsfig]{article}

\def\PRL{{\em Phys. Rev. Lett.}}
\def\PRB{{\em Phys. Rev.} B}
\def\JTPL{{\em JETP Letters}}

\def\be{\begin{equation}}
\def\ee{\end{equation}}
\def\bea{\begin{eqnarray}}
\def\eea{\end{eqnarray}}

\def\Q{{\cal Q}}
\def\W{{\cal W}}

\def\J{{\cal J}}

\newcommand{\Tr}{\mathop{\rm Tr}}
\begin{document}
\title{PROXIMITY EFFECT IN PRESENCE OF QUANTUM FLUCTUATIONS}

\author{M. V. Feigel'man$^1$, A. I. Larkin$^{1,2}$ and M. A. Skvortsov$^1$ }

\address{
$^1$L. D. Landau Institute for Theoretical Physics, Moscow
117940, RUSSIA \\
$^2$Theoretical Physics Institute, University of Minnesota,
Minneapolis, MN 55455, USA
}

\maketitle\abstracts{
The effect of Coulomb interaction upon superconductive
 proximity effect in disordered metals is studied, employing newly
developed Keldysh functional approach. We have calculated subgap Andreev
conductance between superconductor and 2D dirty film, as well as
Josephson coupling via such a film. Both two qualitatively different
Coulomb effects - suppression of the tunneling density of states
and disorder-enchanced repulsion in the Cooper channel - are  shown
to be important at sufficiently low temperatures,
$ \ln^2T\tau \geq  1/g$, where $g=\hbar\sigma/e^2$
 is the dimesionless sheet conductance of the film.
}

\section{Introduction}

Electron transport in hybrid superconductive-normal (S-N) systems at low
temperatures
is governed by the Andreev reflection.
Both finite-voltage conductance $G_A$
between superconductive and normal electrodes, and Josephson critical
current $I_c$ between two superconductive banks, separated by normal
region,  are determined by the Cooper pair propagation in the normal metal.
The theory of Andreev conductance (without Coulomb effects) was developed
in e.~g.~\cite{volkov,Nazarov94,beenakker}, whereas Josephson coupling
was calculated (with the account of short-range electron interaction)
in~\cite{ALO}. When normal conducting region
is made of a dirty metal film, or two-dimensional electron gas with low
density of electrons, Coulomb interaction in the normal region may lead
to strong quantum fluctuations which suppress both Andreev conductance
and Josephson proximity effect.

There are two different quantum effects which play major role in
dirty superconductive systems. First of all,
critical temperature of a uniformly disordered
superconductive film is suppresed by disorder, and eventually
vanishes~\cite{finkel1,finkel2,finkel3}
at the critical value of conductance
$g_c = (2\pi)^{-2}\ln^2\frac1{T_{c0}\tau}$; here $\tau$ is the elastic
scattering time and
 $T_{c0}$ is the BCS transition temperature.
In this range of parameters weak-localization~\cite{band4,GLK}
 and interaction-induced~\cite{AA} corrections to conductivity of 2D
metal are of the
 relative order of $g^{-1}\ln\frac1{T_{c0}\tau} \ll 1$.
 This fact brings about the issue of a quantum (i.e. $T=0$)
 superconductor-metal transition
 (cf.~\cite{Gantm98,finkel3,Spivak95,FeigLar})
 as opposed to the usually assumed~\cite{Mfisher}
direct S-I transition.
The second effect that is important for hybrid
superconductor-normal systems with tunnel barriers at the S/N interface,
is the effect of tunneling conductance suppression
("zero-bias anomaly")
 due to slow relaxation of an extra charge after the tunneling
event~\cite{AA,AAL,finkel2,LS}; as noted in~\cite{LS},
this is physically the same effect as Coulomb blockade of tunneling into a
finite system. In a 2D system with
a long-range Coulomb interaction this effect
scales as
$g^{-1}\ln^2\frac{1}{T\tau}$.
Whereas the first discussed effect
is insensitive to the form of long-range Coulomb asymptotics, the zero-bias
anomaly does crucially depend on it, and may be suppressed by screening
of the Coulomb interaction by nearby external electrodes~\cite{LS}.

In the present paper we discuss behaviour of the subgap conductance
between superconductor and thin dirty metal film, as well as Josephson
coupling between two superconductors via such a film, assuming that
relevant energy scale $\hbar\omega = \max(eV,T)$ is such that
$\sqrt{g} \leq \ln\omega\tau \ll g$.
The second inequality allows us
to neglect weak localization and interaction-induced corrections to the
film conductance, whereas the first one calls for a nonperturbative treatment
of the zero-bias anomaly (ZBA) and Finkelstein
effects~\cite{finkel1,finkel2,finkel3}.
The first of these effects was recently
considered in~\cite{Oreg99}, where renormalization of the
tunneling DoS on the N side of N/S sandwich was calculated within replica
functional method (cf.~\cite{finkel2}). We prefer to avoid the use of
replicas; instead, we have developed
the functional integral approach based on the Keldysh~\cite{Keldysh}
representaion of Green functions
for dirty superconductors~\cite{LO1}.
An advantage of this method is that it allows to calculate
nonequlibrium quantities, and does not contain any analitic
continuation procedures. Recently this method was developed for normal
metals in~\cite{Kamenev98}; in many respects we follow that approach.
Here we just present our main
ideas and results, whereas full details can be found
in forecoming paper~\cite{long}.

\section{Keldysh $\sigma$-model}

In the simplest case of spin-independent interactions, and
neglecting relativistic effects, we can write down an effective
low-energy action for a dirty film of superconductor as~\cite{long}
\bea
 & \displaystyle
  S = \frac{i\pi\nu}{4}
    \Tr \left[
      D (\nabla Q)^2
      + 4i \bigl( i\tau_z \partial_t + {\bf\Phi} + {\bf\Delta} \bigr) Q
    \right]
  + 2\nu \Tr \vec\Phi^T \sigma_x \vec{\Phi}
  + \Tr \vec{\phi}^T \sigma_x \hat{V}_0^{-1} \vec{\phi}
  + \frac{2\nu}{\lambda} \Tr \vec{\Delta}^+ \sigma_x \vec{\Delta}
 &
\nonumber \\
 & \displaystyle
 + \frac{i\pi}{4} \gamma \Tr\nolimits_\Gamma
   e^{i({\bf K}(t)-{\bf K}_S(t))\tau_z} Q
   e^{-i({\bf K}(t)-{\bf K}_S(t))\tau_z} Q_S .
 &
\label{action}
\eea
Here $Q=Q_{t,t'}$ is a $4\times4$ matrix in the $K\otimes N$ space, which
depends upon two time arguments $t,t'$.
 Pauli matrices in the Keldysh ($K$)
$2\times 2$ space are denoted by $\sigma_{x,y,z}$, where $\tau_{x,y,z}$ stand
for the Nambu ($N$) space. Matrix $Q$ is subject to the usual $\sigma$-model
 constraint $Q^2 = {\bf 1}$. Operation $\Tr$ means taking trace over
$K\otimes N$ matrix spaces, as well as over time and real spaces.
 $\nu$ is the bare DoS per one spin
projection at the Fermi level, $D$  is the diffusion coefficient,
 $\phi$ is the (fluctuating)
 electric potential, $\Delta$ is the BCS order parameter field, $\lambda$
is the interaction constant in the Cooper channel (its negative sign corresponds
to attraction). $V_0(q) = 2\pi e^2/q$ is the Fourrier component of Coulomb
interaction in 2D.
We use both vector and tensor (boldfaced) notations for bosonic fields
in the Keldysh space, e.~g.\ $\vec\Delta = (\Delta_1,\Delta_2)^T$ and
${\bf\Delta} = \Delta_1\sigma_0 + \Delta_2\sigma_1$,
where $\Delta_{1,2}$ are certain linear combinations
of the $\Delta$ field on the forward and backward branches
of the Keldysh time integration path.
The field $\vec\Phi = \vec\phi - \partial_t \vec K$, where
$\vec K$ is the phase of the gauge transformation similar to the one proposed
in~\cite{Kamenev98}; in the superconductive case the original matrix
field $\check{Q}$ changes under this gauge transformation according to
$\check{Q}_{t,t'} = e^{i{\bf K}(t)\tau_z}Q_{t,t'}e^{-i{\bf K}(t')\tau_z}$.
Covariant space derivative
$\nabla X\equiv \partial_{\bf r} X + i[\tau_z \partial_{\bf r}{\bf K}, X] $.
The last term in (\ref{action}), denoted below as $S_\gamma$,
describes electron tunneling across the barrier~(cf.~\cite{Oreg99}).
Here $Q_S$ and $K_S$ refer to the S side
of the interface boundary $\Gamma$;
the notation $\Tr\nolimits_\Gamma$ means that the
space integral is taken over the interface surface,
$\gamma$ is the (dimensionless) normal-state tunneling conductance
per unit area of the boundary.
Variation of the total action with respect to $Q$ and $Q_S$ with the
constraints $Q^2=Q_S^2={\bf 1}$ leads (at $K=0$) to the dynamic~\cite{LO1}
 Usadel equation~\cite{Usadel}, together with the standard~\cite{Kupriyanov}
boundary conditions.
In the absence of the boundary term, the equilibrium
saddle-point solution of the Usadel equation is given by
$ Q = u\cdot\mbox{diag}(\hat{Q}^R,\hat{Q}^A)\cdot u$,
where $\hat{Q}^{R(A)}$ are retarded (advanced) Green functions,
$  u = u^{-1} = \sigma_z + \sigma_+ \hat{F}$,
and $\hat{F} = f + f_1\tau_z$ is the generalized fermion distribution
function, $\sigma_+=(\sigma_x+i\sigma_y)/2$.
This representation suggests the use of
a new variable, $\Q = uQu$.

Below we consider low-energy limit $\epsilon \ll |\Delta|$, so the
superconductive matrix reduces to purely phase
rotations: $Q_S = -i {\bf \Delta}/|\Delta|$.
We will also assume weak tunnlling across S/N interface, which makes
it possible to start (on the normal side of interface) from purely
metallic saddle-point $\Q_0=\sigma_z\tau_z$.
Slow rotations of this trivial solution can be parametrized as
$\Q = e^{-{\W}/2}\sigma_z\tau_ze^{{\W}/2}$
by the matrix $\W$ subject to the constraint
$\{\W,\sigma_z\tau_z\} =0$, which is resolved as
$\W = {\scriptstyle \left( \begin{array}{cc}
     w_x \tau_x + w_y \tau_y
     & w_0 + w_z \tau_z \\
     \overline w_0 + \overline w_z \tau_z &
     \overline w_x \tau_x + \overline w_y \tau_y
  \end{array} \right)_K}
$.
The diagonal (off-diagonal) in the Nambu space excitations,
$w_i$ and $\overline{w}_i$ with $i=0,z$ ($i=x,y$),
correspond to diffusion (Cooper) modes in the metal.
Following~\cite{Kamenev98}, we choose ${\bf K}$
to be a linear functional of ${\bf\phi}$
and require the vanishing of the term bilinear in $W$ and
${\bf \Phi}$ in the $\sigma$-model action (\ref{action}).

The term in the action, which is responsible for Andreev subgap
conduction, comes from the averaging of $S_\gamma^2$ over Cooperon
modes: $S_A = \frac{i\pi}{16}G_A\Tr(\Q_S\sigma_z\tau_z)^2$,
where $G_A$ is the (dimensionless) Andreev conductance.
At low energies $(\omega,T,eV) \ll E_{th}= \hbar D/L^2$,
$L$ being the relevant length scale of the N region,
and in the absense of quantum corrections,
the known result~\cite{Nazarov94} $G_A^{(0)} = G_T^2 R_D$ is recovered,
where $R_D$ is the total resistance of the diffusive metal region
and $G_T=\gamma {\cal A}$ is the total normal-state tunneling conductance,
${\cal A}$ being the area of the junction.
In the geometry of rectangular N/S contact
(normal metal film with the length $L_y$ along the boundary with SC,
and the length $L_x$ between SC and normal reserviour),
$G_A^{(0)} = \frac{G_T^2}{g} \frac{L_x}{L_y}$.
At higher $\omega$ and/or $T$, the length $L_x$ should be replaced by
$L_{\rm eff}(\omega,T)$, which is equal to
$L_{\rm eff}(\omega) = \sqrt{2D/|\omega|}$ at $\hbar\omega \gg T$,
and to $L_{\rm eff}(T) = 0.95\sqrt{D/2T}$ in the opposite limit.

\section{Renormalization group}

Quantum fluctuations
lead to renormalization of the Cooper-channel interaction constant
$\lambda$ in the N metal and of the barrier transparancy $\gamma$;
note that the sheet conductance $g = 2\nu D$ is constant within our
approximation.
Renormalization of $\lambda$ comes about in the second
order over the interaction term $S_{\phi,Q} =
-\pi\nu\Tr\left[u{\bf\Phi}u\Q\right] $, which should be averaged over
diffusion/Cooperon modes of the $Q$ matrix field and electric potential
 fluctuations $\phi$. This calculation can be done in the standard gauge
with $K=0$ (cf.~\cite{long} for detailed discussion), the result coincides
with the one known from~\cite{finkel1}:
\be
\frac{d\lambda}{d\zeta} = \frac{1}{4\pi^2 g} -\lambda^2;
 \qquad \zeta = \ln\frac{1}{\omega\tau} ,
\label{lambda}
\ee
where $\omega$ is the running low-frequency cutoff of RG procedure,
which stops eventually at $\omega=\mbox{max}(T,eV,E_{th})$.
The last term in (\ref{lambda}) is the usual BCS ladder contribution, which
can also be considered within RG approach, as coming from the second order
term over $S_{\Delta,Q} = -\pi\nu \Tr\left[u{\bf \Delta}u\Q\right]$.
Eq.~(\ref{lambda}) has a locally stable infrared fixed point
$\lambda_g=1/2\pi\sqrt{g}$.
Integration of Eq.~(\ref{lambda}) with initial condition $\lambda_0 < 0$
 leads to Finkelstein's result for the $T_c$ supression:
$T_c\tau =  \left(\frac{1-\lambda_g\ln(1/T_{c0}\tau)}
{1+\lambda_g\ln(1/T_{c0}\tau)}
\right)^{1/2\lambda_g}$.

There are two sources of renormalization corrections to $\gamma$.
The first one is due to Cooper-channel interaction; to find it one
should average over Cooperon modes the product  $S_\gamma S_\lambda$, where
$S_\lambda =
  (1/4)\pi^2\nu\lambda\int dtd^2{\bf r}
  \,\mbox{tr}\, \sigma_x[Q_{tt}^2-(\tau_z Q_{tt})^2]
$
is the result of Gaussian
integration over $\Delta$ field in the action (\ref{action}).
The second contribution to $\gamma$ is due to fluctuations of the
Coulomb-induced phase ${\bf K}(t)$, describing the ZBA effect; here we
neglect similar fluctuations of the phase ${\bf K}_S(t)$, assuming that
voltage at the superconductor side is fixed by external curcuit, and
SC diffusion constant $D_S \gg D$ (for the opposite situation
cf.\cite{Narozhny99}).
Mean-squared fluctuations of
the field $ K(t)$ can be expressed via the function $\rho(\omega)$
which generalizes a notion of
 "environmental impedance"  introduced in~\cite{Huck97}
within phenomenological approach to the Coulomb suppression of
Andreev conductance. We calculate $\rho(\omega)$
microscopically~\cite{long} for all considered geometries, see below.
The resulting RG equation for $\gamma$ reads
\be
  \frac{d\gamma}{d\zeta} =
  -\gamma\left( \lambda + \frac{2}{\pi}\,\rho\right)
  \quad \mbox{at} \quad
  \zeta > \zeta_{\Delta} = \ln\frac{1}{\Delta\tau}.
\label{gamma}
\ee
At high frequencies $\omega \geq \Delta$ one has instead
$d\gamma/d\zeta = - \gamma\, \rho/\pi$,
as it should be for the tunneling
in the absence of superconductivity~\cite{finkel2,LS,Kamenev98}.
Eqs.~(\ref{lambda}), (\ref{gamma}) are sufficient to study low-energy
behaviour of the Andreev conductance in a rectangular geometry.
Another interesting problem is the subgap conductance
between small SC island of size $d$ and large metal film
(cf.~\cite{FeigLar}),
when the "diffusive" resistance $R_D$ logarithmically depends on the relevant
space scale.
As a result, the Andreev conductance becomes itself subject to the RG
transformation:
\be
  \frac{dG_A}{d\zeta} = {\cal A}^2\frac{\gamma^2}{4\pi g} -
  \frac{4}{\pi}\rho\, G_A \quad \mbox{at} \quad
  \zeta > \zeta_{\Delta},
 \label{GA}
\ee
where we assume, for simplicity,
$\zeta_{\Delta} \approx   \zeta_d = \ln\frac{d^2}{D\tau}$.

\section{Andreev conductance}

We start from the case of rectangular geometry of the contact, when
it suffices to find renormalized value $\gamma_R$
by integration of Eqs.~(\ref{lambda}), (\ref{gamma}), so that
$G_A = G_A^{(0)}(\gamma_R/\gamma_0)^2$.
In the practically interesting (at $\omega \leq \Delta$) limit
$\frac{2\pi\sigma}{\omega} \gg L \gg L_{\rm eff}(\omega)$ we find
$\rho(\omega) = \frac1{\pi g}\ln\frac{L}{L_{\rm eff}(\omega)}$\,
(we assumed here $L=L_x\sim L_y$).
The result for the Andreev conductance reads
\be
\label{Ga-sqr}
  G_A(\omega) =
   \frac{ G_{T\Delta}^2}{g}\,\frac{ L_{\rm eff}(\omega,T)}{L_y} \,
    \frac{4 (\omega/\Delta)^{2\lambda_g} }
    { \left[ 1+ \lambda_n/\lambda_g +
      (1-\lambda_n/\lambda_g) (\omega/\Delta)^{2\lambda_g} \right]^2 }
  \cdot
  \exp \left(
    -\frac{1}{\pi^2 g}\ln\frac{\Delta}{\omega} \cdot
    \ln \frac{\omega\Delta}{E_{th}^2}
  \right)
\ee
where $\lambda_n$ is the Cooper-channel replulsion constant at the energy
scale $\hbar\tau^{-1}$, $G_{T\Delta}$ is the normal-state tunneling
conductance at $eV \approx \Delta$. The last multiplier in the r.h.s. of
Eq.~(\ref{Ga-sqr}) describes the ZBA effect,
with the doubled coefficient in front of the log accounting for the
fact that in this geometry extra charge spreads (after tunneling)
over half-plane.
Eq.~(\ref{Ga-sqr}) is valid at $\omega \geq E_{th} = \frac{\hbar D}{L_x^2}$.
In the opposite limit the result (\ref{Ga-sqr}) can be used after
replacements $\omega\to \frac{D}{L_x^2}$, $L_{\rm eff}(\omega) \to L_x$.

Next we consider geometry of small SC island siting in the middle
of the N thin film of characteristic size $L$.
Now we need to integrate the whole set of Eqs.~(\ref{lambda})--(\ref{GA}).
We assume, for simplicity, that the island size
$d \sim \xi_{\Delta} = \sqrt{D/\Delta}$.
Thus at the energy scale
$\omega \ll \Delta$ extra charge speads symmetrically over 2D plane and
$\rho(\omega) = \frac1{2\pi g}\ln\frac{L}{L_{\rm eff}(\omega)}$.
The result reads
(assuming again $\frac{2\pi\sigma}{\omega} \gg L \gg L_{\rm eff}(\omega)$):
\be
  G_A(\omega) =
  \frac{G_{T\Delta}^2}{4\pi g} \,\frac{1-(\omega/\Delta)^{2\lambda_g}}
{ (\lambda_g + \lambda_n)
+ (\lambda_g - \lambda_n)
  (\omega/\Delta)^{2\lambda_g}}
\cdot
  \exp\left(
    -\frac{1}{2\pi^2 g}\ln\frac{\Delta}{\omega} \cdot
    \ln\frac{\omega\Delta}{E_{th}^2}
  \right)
\label{Ga-isl}
\ee
At lowest energies $\omega$ should be replaced by $E_{th}$
in Eq.~(\ref{Ga-isl}).
The whole effect of quantum fluctuations in this
geometry is that initial "semiclassical" logarithmic growth of $G_A$
with $\omega$ decrease, $G_A \propto \ln\frac{D}{\omega d^2}$,
crosses over to the {\em decrease} by the "log-normal" law.

Note that in both cases (\ref{Ga-sqr}), (\ref{Ga-isl}) the ZBA effect
is represented as a separate multiplicative factor; the reason is that
 fluctuations responsible for the ZBA have 
frequencies 
 $\omega \gg Dq^2$, and thus they are not mixed with
the low-$\omega$ fluctuations respon
sible for the $\lambda$ renormalization.
Detailed dependences of the ZBA factor on $V$ and $T$ can found using
formulae from~\cite{Huck97} with $\rho(\omega)$ employed as the effective
impedance. The power-law factors due to Finkelstein's renormalizations
are determined always by the largest of the scales $eV,T,E_{th}$.

\section{Josephson proximity coupling}

The term in the effective action, which is responsible for the Josepshon
proximity coupling, can be written in the form
$
S_J = \frac{1}{2}E_J \Tr[\check{Q}_S^{(1)}\check{Q}_S^{(2)}\sigma_x]
$,
where superscrits $^{(1)}$ and $^{(2)}$ refer to two superconductive
banks or islands. Using low-energy representation for $Q_S$,
and neglecting phase factors $\exp(i{\bf K}(t)\tau_z)$,
one finds that calculation of the Josephson current with the use of $S_J$ 
produces standard expression
$ I_J =  \frac{2e}{\hbar}E_J\sin(\theta_1-\theta_2)$.
The contribution of $K(t)$ fluctuations factorizes as above,
and is taken care of by the same multilicative factor as in
Eqs.~(\ref{Ga-sqr}), (\ref{Ga-isl}),
with $\omega$ replaced by an appropriate inverse diffusion time.
We start from the calculation of $\tilde{E}_J$ in the absense of the ZBA.
For both geometries discussed above,
$\tilde{E}_J$ can be expressed via the same function defined
in the Fourier space,
the zero-frequency Cooperon amplitude $\J(q)$, which obeys the RG
equation
$
  \frac{\partial\J(\zeta_q)}{\partial\zeta_q} =
  {\cal A}^2 \frac{\gamma^2(\zeta_q)}{8\nu}
$,
where $\zeta_q =
\ln(\Delta/Dq^2)$.
First we consider an example of two small SC islands of radius $d$,
separated by the distance $R \gg d$.
In this case $\tilde{E}_J$, which is given by the inverse
Fourrier transform of $\J(q)$, can be represented as
$
  \tilde{E}_J(R)
  = \frac{1}{\pi R^2}
      \left. \frac{\partial\J(\zeta_q)}{\partial\zeta_q}
      \right|_{\zeta_q = \ln\frac{R^2}{D\tau}}
$.
Using renormalized $\gamma(\zeta_q)$ according to
Eqs.(\ref{lambda},\ref{gamma}), and adding the ZBA factor from
(\ref{Ga-isl}) estimated at $\omega = D/R^2$, we find (with 
the total size $L$ of the film being in the range 
$R \ll L \ll 4\pi\nu e^2 R^2$ ):
\be
E_J(R) = \frac{\lambda_g^2\, G_{T\Delta}^2}{2\pi\nu ( \lambda_n +\lambda_g)^2 }
\frac{1}{R^2}
  \left(\frac{\xi_{\Delta}}{R}\right)^{4\lambda_g}
\frac{1}{\left[1+ \beta(\xi_{\Delta} /R)^{4\lambda_g}\right]^2}
\exp\left(-\frac{2}{\pi^2 g}\ln\frac{R}{\xi_{\Delta}} 
\ln\frac{L^2}{\xi_{\Delta} R}\right)
\label{EJ1}
\ee
where $\beta =\frac{\lambda_g-\lambda_n}{\lambda_g +\lambda_n}$.
The increase of the power-law exponent up to $x_J = 2 + 4\lambda_g$ is
due to Finkelstein's corrections, whereas
additional log-normal decay factor is due to the ZBA.

In the case of rectangular geometry the discrete nature of diffusion
modes in the N region should be taken into account~\cite{long} while
calculating $E_J$.
The result  (for $L_x \sim L_y$ ) is:
\be
E_J(L_x,L_y) =
\frac{\lambda_g\, G_{T\Delta}^2}{4\nu (\lambda_n + \lambda_g)^2 }\frac{1}{L_xL_y}
  \left(\frac{\xi_{\Delta}}{L_x}\right)^{4\lambda_g}
\frac{1}{1+ \beta(\xi_{\Delta}/L_x)^{4\lambda_g}}
\exp\left(-\frac{4}{\pi^2 g}\ln^2\frac{L_x}{\xi_{\Delta}}\right)
\label{EJ2}
\ee

This research was started during the ICTP Workshop "Disorder and chaos
in mesoscopic systems", Trieste, August 1998.
We are grateful to I.L.Aleiner, A.Kamenev, A.S.Ioselevich, G.B.Lesovik, 
Yu.V.Nazarov, B.Z.Spivak and F.Zhou for useful discussions.
We thank financial support from the NSF grant DMR-9812340, 
DGA grant \# 94-1189,
Swiss NSF grant \# 7SUP J048531, and
RFBR grant \# 98-02-19252.

\end{document}